
!!!!!!!!!!!!!!!!!!!!!!!!!!!!!!!!!!!!!!!!!!!!!!!!!!!!!!!!!!!!!!!!!!!!!!
%
\newif\ifproofmode			
\proofmodefalse				

\newif\ifforwardreference		
\forwardreferencetrue			

\newif\ifchapternumbers			
\chapternumbersfalse			

\newif\ifcontinuousnumbering		
\continuousnumberingfalse		

\newif\iffigurechapternumbers		
\figurechapternumbersfalse		

\newif\ifcontinuousfigurenumbering	
\continuousfigurenumberingfalse		

\newif\ifcontinuousreferencenumbering   
\continuousreferencenumberingtrue       

\newif\ifparenequations			
\parenequationstrue			

\newif\ifstillreading			

\font\eqsixrm=CMR6			
\def\marginstyle{\eqsixrm}		

\newtoks\chapletter			
\newcount\chapno			
\newcount\eqlabelno			
\newcount\figureno			
\newcount\referenceno			
\newcount\minutes			
\newcount\hours				

\newread\labelfile			
\newwrite\labelfileout			
\newwrite\allcrossfile			

\chapno=0
\eqlabelno=0
\figureno=0

%
\def\initialeqmacro{
    \ifproofmode
        \headline{\tenrm \today\ --\ \timeofday\hfill
                         \jobname\ --- draft\hfill\folio}
        \hoffset=-1cm
        \immediate\openout\allcrossfile=zallcrossreferfile
    \fi
    \ifforwardreference
        \openin\labelfile=zlabelfile
        \ifeof\labelfile
        \else
            \stillreadingtrue
            \loop
                \read\labelfile to \nextline
                \ifeof\labelfile
                    \stillreadingfalse
                \else
                    \nextline
                \fi
                \ifstillreading
            \repeat
        \fi
        \ifproofmode
            \immediate\openout\labelfileout=zlabelfile
        \fi
    \fi}


{\catcode`\^^I=9
\catcode`\ =9
\catcode`\^^M=9

%
\gdef\chapfolio{			
    \ifnum \chapno>0 \relax		
        \the\chapno
    \else
        \the\chapletter
    \fi}

%
\gdef\bumpchapno{
    \ifnum \chapno>-1 \relax
        \global \advance \chapno by 1
    \else
        \global \advance \chapno by -1 \setletter\chapno
    \fi
    \ifcontinuousnumbering
    \else
        \global\eqlabelno=0
    \fi
    \ifcontinuousfigurenumbering
    \else
        \global\figureno=0
    \fi
    \ifcontinuousreferencenumbering
    \else
        \global\referenceno=0
    \fi}

%
\gdef\setletter#1{\ifcase-#1 {}  \or\global\chapletter={A}
  \or\global\chapletter={B} \or\global\chapletter={C} \or\global\chapletter={D}
  \or\global\chapletter={E} \or\global\chapletter={F} \or\global\chapletter={G}
  \or\global\chapletter={H} \or\global\chapletter={I} \or\global\chapletter={J}
  \or\global\chapletter={K} \or\global\chapletter={L} \or\global\chapletter={M}
  \or\global\chapletter={N} \or\global\chapletter={O} \or\global\chapletter={P}
  \or\global\chapletter={Q} \or\global\chapletter={R} \or\global\chapletter={S}
  \or\global\chapletter={T} \or\global\chapletter={U} \or\global\chapletter={V}
  \or\global\chapletter={W} \or\global\chapletter={X} \or\global\chapletter={Y}
  \or\global\chapletter={Z}\fi}

%
\gdef\tempsetletter#1{\ifcase-#1 {}\or{} \or\chapletter={A} \or\chapletter={B}
 \or\chapletter={C} \or\chapletter={D} \or\chapletter={E}
  \or\chapletter={F} \or\chapletter={G} \or\chapletter={H}
   \or\chapletter={I} \or\chapletter={J} \or\chapletter={K}
    \or\chapletter={L} \or\chapletter={M} \or\chapletter={N}
     \or\chapletter={O} \or\chapletter={P} \or\chapletter={Q}
      \or\chapletter={R} \or\chapletter={S} \or\chapletter={T}
       \or\chapletter={U} \or\chapletter={V} \or\chapletter={W}
        \or\chapletter={X} \or\chapletter={Y} \or\chapletter={Z}\fi}

%
\gdef\chapshow#1{
    \ifnum #1>0 \relax
        #1
    \else
        {\tempsetletter{\number#1}\the\chapletter}
    \fi}

%
\gdef\today{\number\day\space \ifcase\month\or Jan\or Feb\or
        Mar\or Apr\or May\or Jun\or Jul\or Aug\or Sep\or
        Oct\or Nov\or Dec\fi, \space\number\year}

\gdef\timeofday{\minutes=\time    \hours=\time
        \divide \hours by 60
        \multiply \hours by 60
        \advance \minutes by -\hours
        \divide \hours by 60
        \ifnum\the\minutes>9
		\the\hours:\the\minutes
 	\else
  		\the\hours:0\the\minutes
	\fi}


%
%
\gdef\chapnum{\bumpchapno \chapfolio}			

\gdef\chaplabel#1{                                      
    \bumpchapno                                         
    \ifproofmode                                        
        \ifforwardreference                             
            \immediate\write\labelfile{
            \noexpand\expandafter\noexpand\def
            \noexpand\csname CHAPLABEL#1\endcsname{\the\chapno}}
        \fi
    \fi
    \global\expandafter\edef\csname CHAPLABEL#1\endcsname
    {\the\chapno}	 				
    \ifproofmode
        \llap{\hbox{\marginstyle #1\ }}
    \fi
    \chapfolio}

%
%
\gdef\eqnum{                                    
    \global\advance\eqlabelno by 1              
    \eqno(
    \ifchapternumbers
        \chapfolio.
    \fi
    \the\eqlabelno)}

\gdef\eqlabel#1{                                
    \global\advance\eqlabelno by 1              
    \ifproofmode                                
        \ifforwardreference
            \immediate\write\labelfileout{\noexpand\expandafter\noexpand\def
            \noexpand\csname EQLABEL#1\endcsname
            {\the\chapno.\the\eqlabelno?}}
        \fi
    \fi
    \global\expandafter\edef\csname EQLABEL#1\endcsname
    {\the\chapno.\the\eqlabelno?}
    \eqno(
    \ifchapternumbers
        \chapfolio.
    \fi
    \the\eqlabelno)
    \ifproofmode
        \rlap{\hbox{\marginstyle #1}}		
    \fi}

\gdef\eqalignnum{                               
    \global\advance\eqlabelno by 1              
    &(\ifchapternumbers
        \chapfolio.
    \fi
    \the\eqlabelno)}

\gdef\eqalignlabel#1{                   	
    \global\advance\eqlabelno by 1 	        
    \ifproofmode		  		
        \ifforwardreference
            \immediate\write\labelfileout{\noexpand\expandafter\noexpand\def
            \noexpand\csname EQLABEL#1\endcsname
            {\the\chapno.\the\eqlabelno?}}
        \fi
    \fi
    \global\expandafter\edef\csname EQLABEL#1\endcsname
    {\the\chapno.\the\eqlabelno?}
    &(\ifchapternumbers
        \chapfolio.
    \fi
    \the\eqlabelno)
    \ifproofmode
        \rlap{\hbox{\marginstyle #1}}			
    \fi}

\gdef\eqref#1{\ifparenequations(\fi
    \ifundefined{EQLABEL#1}***
        \ifproofmode
            \ifforwardreference)
            \else
                \write16{ ***Undefined\space Equation\space Reference #1*** }
            \fi
        \else
            \write16{ ***Undefined\space Equation\space Reference #1*** }
        \fi
    \else
        \edef\LABxx{\getlabel{EQLABEL#1}}
        \def\LAByy{\expandafter\stripchap\LABxx}
        \ifchapternumbers
            \chapshow{\LAByy}.\expandafter\stripeq\LABxx
        \else
            \ifnum \number\LAByy=\chapno \relax
                \expandafter\stripeq\LABxx
            \else
                \chapshow{\LAByy}.\expandafter\stripeq\LABxx
            \fi
        \fi
        \ifparenequations)\fi
    \fi
    \ifproofmode
        \write\allcrossfile{Equation #1}
    \fi}

%
\gdef\fignum{                                   
    \global\advance\figureno by 1\relax         
    \iffigurechapternumbers
        \chapfolio.
    \fi
    \the\figureno}

\gdef\figlabel#1{				
    \global\advance\figureno by 1\relax 	
    \ifproofmode				
        \ifforwardreference
            \immediate\write\labelfileout{\noexpand\expandafter\noexpand\def
            \noexpand\csname FIGLABEL#1\endcsname
            {\the\chapno.\the\figureno?}}
        \fi
    \fi
    \global\expandafter\edef\csname FIGLABEL#1\endcsname
    {\the\chapno.\the\figureno?}
    \iffigurechapternumbers
        \chapfolio.
    \fi
    \ifproofmode
        \llap{\hbox{\marginstyle #1\ }}\relax
    \fi
    \the\figureno}

\gdef\figref#1{					
    \ifundefined				
        {FIGLABEL#1}!!!!			
        \ifproofmode
            \ifforwardreference
            \else
                \write16{ ***Undefined\space Figure\space Reference #1*** }
            \fi
        \else
            \write16{ ***Undefined\space Figure\space Reference #1*** }
        \fi
    \else
        \edef\LABxx{\getlabel{FIGLABEL#1}}
        \def\LAByy{\expandafter\stripchap\LABxx}
        \iffigurechapternumbers
            \chapshow{\LAByy}.\expandafter\stripeq\LABxx
        \else \ifnum\number\LAByy=\chapno \relax
                \expandafter\stripeq\LABxx
            \else
                \chapshow{\LAByy}.\expandafter\stripeq\LABxx
            \fi
        \fi
        \ifproofmode
            \write\allcrossfile{Figure #1}
        \fi
    \fi}

%
\gdef\refnum{                                      
    \global\advance\referenceno by 1\relax         
    \the\figureno}	                           

\gdef\internalreflabel#1{			
    \global\advance\referenceno by 1\relax 	
    \ifproofmode	   			
        \ifforwardreference
            \immediate\write\labelfileout{\noexpand\expandafter\noexpand\def
            \noexpand\csname REFLABEL#1\endcsname
            {\the\chapno.\the\referenceno?}}
        \fi
    \fi
    \global\expandafter\edef\csname REFLABEL#1\endcsname
    {\the\chapno.\the\figureno?}
    \ifproofmode
        \llap{\hbox{\marginstyle #1\hskip.5cm}}\relax
    \fi
    \the\referenceno}

\gdef\internalrefref#1{				
    \ifundefined				
        {REFLABEL#1}!!!!			
        \ifproofmode
            \ifforwardreference
            \else
                \write16{ ***Undefined\space Footnote\space Reference #1*** }
            \fi
        \else
            \write16{ ***Undefined\space Footnote\space Reference #1*** }
        \fi
    \else
        \edef\LABxx{\getlabel{REFLABEL#1}}
        \def\LAByy{\expandafter\stripchap\LABxx}
        \expandafter\stripeq\LABxx
        \ifproofmode
            \write\allcrossfile{Figure #1}
        \fi
    \fi}

%
\gdef\reflabel#1{\item{\internalreflabel{#1}.}}

%
\gdef\refref#1{\internalrefref{#1}}

\gdef\eq{\ifhmode Eq.~\else Equation~\fi}		
\gdef\eqs{\ifhmode Eqs.~\else Equations~\fi}

%
%

%
\gdef\getlabel#1{\csname#1\endcsname}
\gdef\ifundefined#1{\expandafter\ifx\csname#1\endcsname\relax}
\gdef\stripchap#1.#2?{#1}				
\gdef\stripeq#1.#2?{#2}					
}  

\overfullrule = 0pt
\magnification = 1200
\baselineskip 21pt plus 0.2 pt minus 0.2 pt
\chapternumberstrue
\forwardreferencetrue
\initialeqmacro
\def\sh{\mathop{\rm sh}\nolimits}
\def\ch{\mathop{\rm ch}\nolimits}

\def\tr{\mathop{\rm tr}\nolimits}
\font\zch=PSMZCHMI scaled 1000
\def\g{\hbox{\zch g}}
\newbox\lett
\newdimen\lheight
\newdimen\lwidth
\def\ontop#1#2{\setbox\lett=\hbox{#2}%
   \lheight\ht\lett \multiply\lheight by 12 \divide\lheight by 10\relax%
   \lwidth\wd\lett \multiply\lwidth by 8 \divide\lwidth by 10\relax%
   #2\kern-\lwidth%
       \raise\lheight\hbox{{$\scriptstyle #1$}}\kern.1ex}

\line{\hfill UMTG-166}

\vskip 0.2 in

\centerline{\bf Analytical Bethe Ansatz}
\centerline{\bf for Quantum-Algebra-Invariant Spin Chains}
\bigskip

\medskip

\centerline{Luca Mezincescu and Rafael I. Nepomechie}
\centerline{Department of Physics}
\centerline{University of Miami, Coral Gables, FL 33124, USA}

\vskip 0.2 in

\bigskip

\centerline{\bf Abstract}

\vskip 0.2 in

We have recently constructed a large class of open quantum spin chains
which have quantum-algebra symmetry and which are integrable.
We show here that these models can be exactly solved using a generalization
of the analytical
Bethe Ansatz (BA) method. In particular, we determine in this way the spectrum
of the transfer matrices of the $U_q [(su(2)]$-invariant spin chains associated
with $A^{(1)}_1$ and $A^{(2)}_2$ in the fundamental representation. The
quantum-algebra invariance of these models plays an essential role in
obtaining these results. The BA equations for these open chains
are ``doubled'' with respect to the BA equations for the corresponding
closed chains.

\vfill\eject

\noindent
{\bf \chapnum . Introduction}
\vskip 0.2truein

Consider the open quantum spin chain consisting of $N$ spins with Hamiltonian
$$\eqalignno{
H &= \sum_{k=1}^{N-1} \Bigl\{ \sigma^1_k \sigma^1_{k+1} +
\sigma^2_k \sigma^2_{k+1} + {1\over 2}( q + q^{-1}) \sigma^3_k \sigma^3_{k+1}
\Bigr\} \cr
&\quad\quad\quad\quad\quad - {1\over 2}( q - q^{-1})
\Bigl( \sigma^3_1 - \sigma^3_N \Bigr) \,, \eqalignlabel{xxz} \cr}
$$
where $\vec\sigma$ are the usual Pauli matrices, and $q$ is an arbitrary
complex parameter. This model is${}^{\refref{alcaraz}, \refref{sklyanin}}$
integrable. Moreover, $H$
commutes${}^{\refref{pasquier},\refref{kulish/sklyanin(1)}}$ with the
generators $S^3$, $S^\pm$ of the quantum algebra $U_q[su(2)]$,
$$ \left[ S^+ \,, S^-\right] = {q^{2S^3} - q^{-2S^3}\over q - q^{-1}}
\,, \quad\quad\quad \left[ S^3 \,, S^\pm \right] = \pm S^\pm \,,
\eqlabel{algebra}  $$
where
$$S^3 = \sum_{k=1}^N S^3_k \,, \quad\quad
S^\pm = \sum_{k=1}^N q^{(S^3_N + \cdots + S^3_{k+1})}\ S^\pm_k \
q^{-(S^3_{k-1} + \cdots + S^3_1)} \,, \eqlabel{comult}  $$
and
$$S^3_k = {1\over 2}\sigma^3_k \,, \quad\quad\quad
S^\pm_k = {1\over 2}(\sigma^1_k \pm i \sigma^2_k) \,. \eqlabel{spins(1)} $$
For $|q| = 1$, this model is critical and is related${}^{\refref{alcaraz},
\refref{pasquier}}$ to the $c < 1$ minimal models.

A pertinent question is whether these results can be generalized to spins
in higher-dimensional representations and to larger symmetry algebras. The
key is to formulate${}^{\refref{sklyanin},\refref{jpa}}$ open spin chains
in terms of so-called $R$ and $K$ matrices which are associated with affine
algebras $g^{(k)}$, where $g$ is a simple Lie algebra ($A_n$ = $su(n+1)$,
$B_n$ = $o(2n+1)$, $C_n$ = $sp(2n)$, $D_n$ = $o(2n)$, etc.) and
$k$ ($=1,2,3$) is the order of a diagram automorphism of $g$.
Following this approach, we have recently constructed${}^{\refref{ijmpa},
\refref{mpla}}$ open quantum spin chains associated with
$A^{(1)}_1$, $A^{(2)}_{2n}$, $A^{(2)}_{2n - 1}$,
$B^{(1)}_{n}$, $C^{(1)}_{n}$ and $D^{(1)}_{n}$
in the fundamental representation. These chains are integrable, and have
the quantum-algebra invariance $U_q[g_0]$, where $g_0$ is the maximal
finite-dimensional subalgebra of $g^{(k)}$. The simplest case of $A^{(1)}_1$
corresponds to the model \eqref{xxz}. (The corresponding chains with spins
in higher-dimensional representations can be constructed using a fusion
procedure${}^{\refref{fusion}-\refref{npb}}$.) A subset of these models, in
a different formulation, has been discussed independently in
Ref. \refref{batchelor}.

While the open $A^{(1)}_1$ chain has been solved both by the
coordinate${}^{\refref{alcaraz}}$ Bethe Ansatz (BA) and the
algebraic${}^{\refref{sklyanin}}$ BA, the new models with
larger symmetry algebras have remained unsolved until now.

In this paper, we show that these models can be exactly solved using
a generalization of the analytical Bethe Ansatz method.
In particular, we determine in this way
the spectrum of the transfer matrices of the $U_q[su(2)]$-invariant
spin chains associated with $A^{(1)}_1$ and $A^{(2)}_2$ in the
fundamental representation. The quantum-algebra invariance of these
models plays an essential role in obtaining these results. We find
that the BA equations for these open chains are ``doubled'' with
respect to the BA equations for the corresponding closed chains.

The analytical BA method, which is a generalization of the ``inverse
transfer matrix'' method${}^{\refref{baxter} -\refref{zamolodchikov}}$,
was developed by
Reshetikhin${}^{\refref{reshetikhin}}$ to determine the spectrum of
transfer matrices for {\it closed} chains associated with affine
algebras $g^{(k)}$. In this approach, one uses general properties of
the $R$ matrix (such as analyticity, unitarity, crossing symmetry, etc.)
to derive various properties of the transfer-matrix eigenvalues.
These properties are used to completely determine the eigenvalues,
assuming that they have the form of ``dressed'' pseudovacuum eigenvalues.
We remark that yet another approach for solving closed chains is the
nested${}^{\refref{nested}}$ BA method. This method is able to produce
also the eigenvectors of the transfer matrix, which is beyond the scope
of the analytical BA method in its present state of development.

The outline of this paper is as follows. In Section 2, we describe the
class of models which we investigate. In particular, we
recall${}^{\refref{ijmpa},\refref{mpla}}$ the expression for the transfer
matrix, which commutes with generators of a quantum algebra. In Section 3,
we describe five general properties of the eigenvalues of the transfer matrix:
fusion property${}^{\refref{npb}}$, crossing symmetry, asymptotic behavior,
periodicity, and analyticity.
We use these properties, together with an appropriate Ansatz,
to determine the eigenvalues for the cases
$A^{(1)}_1$ and $A^{(2)}_2$ in Sections 4 and 5, respectively. Section 6
contains some concluding remarks. There are three appendices.

\vskip 0.4truein

\noindent
{\bf \chapnum .  The model}

\vskip 0.2truein

We consider the open quantum spin chain with
Hamiltonian${}^{\refref{ijmpa},\refref{mpla}}$
$$H = \sum_{k=1}^{N-1} {d\over du} \check R_{k,k+1}(u) \Big\vert_{u=0}
\,, \eqlabel{hamiltonian} $$
whose state space is the $N$-fold tensor product
$C^n\otimes \cdots \otimes C^n$.
The notations here are standard. (See, e.g., Refs. \refref{kulish/sklyanin(2)}
and \refref{kulish/sklyanin(3)}.)
In particular, $\check R(u) = {\cal P} R(u)$, where
${\cal P}$ is the permutation matrix in $C^n \otimes C^n$
(i.e., ${\cal P} (x \otimes y) = y \otimes x$ for $x, y \in C^n$),
and $R(u)$ is a matrix acting in $C^n \otimes C^n$
which obeys the Yang-Baxter equation
$$
R_{12}(u - v)\  R_{13}(u)\ R_{23}(v) = R_{23}(v)\  R_{13}(u)\ R_{12}(u - v)
\,.   \eqlabel{yang-baxter}
$$
As usual, $R_{12}(u)$, $R_{13}(u)$ and $R_{23}(u)$ are matrices acting
in $C^n \otimes C^n \otimes C^n$, with $R_{12}(u) = R(u) \otimes 1$,
$R_{23}(u) = 1 \otimes R(u)$, etc.

We restrict to the $R$ matrices
associated with $A^{(1)}_1$, $A^{(2)}_{2n}$, $A^{(2)}_{2n - 1}$,
$B^{(1)}_{n}$, $C^{(1)}_{n}$ and $D^{(1)}_{n}$
in the fundamental representation, which were found
by Bazhanov${}^{\refref{bazhanov}}$ and Jimbo${}^{\refref{jimbo}}$.
These matrices satisfy the following additional properties,
which we shall extensively exploit:
$PT$ symmetry
$${\cal P}_{12}\ R_{12}(u)\ {\cal P}_{12} \equiv R_{21}(u)
= R_{12}(u)^{t_1 t_2} \,,
\eqlabel{pt}
$$
where $t_i$ denotes transposition in the $i^{th}$ space; unitarity
$$R_{12}(u)\  R_{21}(-u) = \zeta (u) \,, \eqlabel{unitarity} $$
where $\zeta (u)$ is some even scalar function of $u$; crossing symmetry
$$R_{12}(u) = V_1 \ R_{12}(-u - \rho)^{t_2}\  V_1
=  V_2^{t_2} \ R_{12}(-u - \rho)^{t_1}\  V_2^{t_2}  \,,
\eqlabel{crossing} $$
where $V^2 = 1$ (in this paper, we use the notation $V_1 = V \otimes 1$,
$V_2 = 1 \otimes V$); regularity
$$R_{12}(0) = \zeta(0)^{1/2}\ {\cal P}_{12} \,; \eqlabel{regularity} $$
and (in the so-called homogeneous gauge used by Jimbo) commutativity
$$\left[ \check R_{12}(u) \,, \check R_{12}(v) \right] = 0  \,.
\eqlabel{Rcheck}  $$
Finally, the matrices $R(u)$ have certain asymptotic behavior for large $u$,
as well as certain periodicity and analyticity properties, about which we
shall say more later.

The transfer matrix corresponding to this Hamiltonian
is${}^{\refref{ijmpa},\refref{mpla}}$
$$t(u) = \tr_a M_a\ T_a(u)\  \hat T_a(u) \,, \eqlabel{transfer} $$
where
$$M = V^t\ V = M^t \,, \eqnum  $$
and
$$\eqalignno{
T_a(u) &= R_{aN}(u)\ R_{a N-1}(u)\ \cdots R_{a1}(u) \,, \cr
\hat T_a(u) &= R_{1a}(u)\ \cdots R_{N-1 a}(u)\ R_{Na}(u) \,.
\eqalignlabel{monodromy} \cr} $$
The subscript $a$ denotes the auxiliary space. (As usual, we suppress the
quantum-space subscripts $1 \,, \cdots \,, N$ of $T_a(u)$ and $\hat T_a(u)$.)
The transfer matrix is also equal to
$$t(u) = \tr_a M_a^{-1}\ \hat T_a(u)\ T_a(u)
\,. \eqlabel{transfertoo} $$
This fact, which follows from an identity which is proved in Appendix A,
will be used in deriving a number of results in this paper.
One can show${}^{\refref{sklyanin},\refref{ijmpa}}$ that
$t(u)$ constitutes a one-parameter commutative family
$$\left[ t(u) \,, t(v) \right] = 0 \hbox{   for all   } u \,, v
\,, \eqlabel{commutativity} $$
which is related to the Hamiltonian \eqref{hamiltonian} by
$$H = {1\over 2 \zeta(0)^{N - {1\over 2}}\left( \tr_a M_a \right)}
{d\over du} t(u) \Big\vert_{u=0} \,.
\eqnum $$

For $R$ matrices associated with the affine algebra $g^{(k)}$,
the leading asymptotic behavior of $T_a(u)$ and $\hat T_a(u)$ for
large $u$ is given by
$$T_a(u) \sim \kappa(u)\ T^+_a \,, \quad\quad\quad
\hat T_a(u) \sim \kappa(u)\ \hat T^+_a \quad\quad {\hbox{  for  }}
u \rightarrow \infty\,.  \eqlabel{asymptoticT} $$
Here $\kappa(u)$ is a scalar function of $u$, and
$T^+_a$ and $\hat T^+_a$ are $u$-independent
triangular matrices which can be
expressed in terms of generators of the quantum algebra $U_q[g_0]$,
where $g_0$ is the maximal finite-dimensional subalgebra of $g^{(k)}$.
The diagonal entries of $T^+_a$ and $\hat T^+_a$ involve the Cartan
generators, while the off-diagonal elements are either raising or lowering
operators. (See, e.g., Ref. \refref{faddeev}.)

Following Kulish and Sklyanin${}^{\refref{kulish/sklyanin(1)}}$, one
can show${}^{\refref{mpla}}$ that the transfer matrix $t(u)$ commutes
with these operators,
$$\left[ t(u) \,,  T^+_a  \right] = 0 \,, \quad\quad\quad
  \left[ t(u) \,,  \hat T^+_a  \right] = 0 \,. \eqlabel{commute}  $$
It follows that
$$ \left[ t(u) \,, U_q [g_0] \right] = 0 \,. \eqlabel{qinvariance}  $$
That is, not only the Hamiltonian \eqref{hamiltonian} but also the
transfer matrix \eqref{transfer} commutes with generators of a quantum
algebra. (For a more detailed review, see Ref. \refref{miami}.)

Let $\left\{ H_1 \,, \cdots \,, H_l \right\}$ be the (Hermitian) generators
of the Cartan subalgebra of $U_q[g_0]$. If the $R$ matrix is real, then
the transfer matrix $t(u)$ is Hermitian. (See Appendix B.) It follows that
there
exist simultaneous eigenstates $| \Lambda^{(h_1 \,, \cdots \,, h_l)} >$,
$$\eqalignno{
t(u)\ | \Lambda^{(h_1 \,, \cdots \,, h_l)} > &=
\Lambda^{(h_1 \,, \cdots \,, h_l)}(u)\
| \Lambda^{(h_1 \,, \cdots \,, h_l)} > \,, \cr
 H_i\ | \Lambda^{(h_1 \,, \cdots \,, h_l)} > &=  h_i\
| \Lambda^{(h_1 \,, \cdots \,, h_l)} > \,, \quad\quad i = 1 \,, \cdots \,, l
\,. \eqalignnum \cr}$$
Our task is to determine $\Lambda^{(h_1 \,, \cdots \,, h_l)}(u)$, the spectrum
of the transfer matrix $t(u)$. In the next Section, we describe general
properties of these eigenvalues. Taking an appropriate Ansatz for the general
form of the eigenvalues, these properties are sufficient to completely
determine them.

\vskip 0.4truein
\noindent
{\bf \chapnum . Properties of Transfer-Matrix Eigenvalues}
\vskip 0.2truein

Here we describe five general properties of the eigenvalues of the transfer
matrix, which follow from properties of the corresponding $R$ matrix.

\medskip
\noindent
{\it Fusion property}
\medskip

In a previous paper${}^{\refref{npb}}$,
we describe a fusion procedure for open chains. By fusing in the auxiliary
space, we construct a ``fused'' transfer matrix $\tilde t(u)$. In order to
explain this construction in more detail, we first
recall${}^{\refref{sklyanin},\refref{jpa}}$ that the transfer matrix
for a very large class of integrable open chains is given by
$$t(u) = \tr_a K^+_a(u)\ T_a(u)\ K^-_a(u)\ \hat T_a(u) \,.  \eqnum $$
The auxiliary-space
matrices $K^-(u)$ and $K^+(u)$, which can be interpreted as amplitudes
for elastic reflection from walls, obey certain relations which ensure that
reflections are consistent with factorized scattering.
The transfer matrix \eqref{transfer}
corresponds to the special case${}^{\refref{ijmpa}, \refref{mpla}}$
$$K^-(u) = 1 \,, \quad\quad\quad K^+(u)= M \,, \eqnum  $$
for which quantum-algebra invariance is achieved.

The corresponding
fused transfer matrix $\tilde t(u)$ is constructed from the
fused $R$ matrices $R_{<12>3}(u)$ and $R_{3 <12>}(u)$, which are given by
$$\eqalignno{
R_{<12> 3}(u) &= {1\over \alpha(u)} \tilde P_{12}^+\ R_{13}(u)\
R_{23}(u + \rho)\ \tilde P_{12}^+  \,, \cr
R_{3 <12>}(u + \rho) &= {1\over \alpha(u)}
\tilde P_{12}^+\ R_{32}(u)\ R_{31}(u + \rho)\
\tilde P_{12}^+  \,, \eqalignlabel{fusedR} \cr} $$
and the fused $K$ matrices $K^\mp_{<12>}(u)$, which are given by
$$\eqalignno{
K^-_{<12>}(u) &= {1\over \beta(u)} \tilde P_{12}^+\ R_{21} (2u + \rho)\
\tilde P_{21}^+  \,, \cr
K^+_{<12>}(u) &= {1\over \beta(u)}
\tilde P_{21}^+\ R_{12} (-2u -3\rho)\ M_1\
M_2\ \tilde P_{12}^+  \,. \eqalignlabel{fusedK} \cr} $$
The projection operator
$\tilde P_{12}^+ =  \left( \tilde P_{12}^+ \right)^2 $ is given by
$$\tilde P_{12}^+ = 1 - \tilde P_{12}^- \,, \quad\quad\quad
\tilde P_{12}^- = {1\over n\ \zeta(0)^{1/2}} R_{12}(-\rho) =
{1\over n} V_1\ {\cal P}_{12}^{t_2}\ V_1 \,, \eqlabel{projector} $$
and
$\tilde P_{21}^+ = \left( \tilde P_{12}^+ \right)^{t_1 t_2}$.
The numerators in the above expressions for the fused $R$ and $K$ matrices are
equal to zero at $u=-\rho$. We remove these ``kinematic zeroes'' by
introducing scalar functions $\alpha(u)$ and $\beta(u)$ in the denominators
which also vanish at $u=-\rho$, such that the fused $R$ and $K$ matrices
are regular (i.e., have no poles) at this value of $u$.

The fused transfer matrix
$\tilde t(u)$ and the original transfer matrix $t(u)$ are related by the
fusion formula${}^{\refref{npb}}$
$$\tilde t(u) = {1\over \alpha(u)^{2N}\ \beta(u)^2}\left\{
\zeta( 2u + 2\rho)\ t(u)\ t(u+\rho) -
\zeta( u + \rho)^{2N}\  \g(2u + \rho)\ \g(-2u - 3\rho) \right\}
\,, \eqlabel{fusion-transfer} $$
where the scalar function $\g(u)$ is defined by
$$\g(u) = \tr_{12}\ R_{12}(u)\ V_1\ V_2\ \tilde P_{12}^-
\,, \eqlabel{calg} $$
and satisfies
$$ \g(u)\ \g(-u) = \zeta(u)  \,. \eqnum $$
Acting with the fusion formula on an eigenstate $|\Lambda>$ of $t(u)$,
we obtain the relation
$$\tilde \Lambda(u) = {1\over \alpha(u)^{2N}\ \beta(u)^2}\left\{
\zeta( 2u + 2\rho)\ \Lambda(u)\ \Lambda(u+\rho) -
\zeta( u + \rho)^{2N}\  \g(2u + \rho)\ \g(-2u - 3\rho) \right\}
\,, \eqlabel{fusion-lambda} $$
which, as we shall see, can be regarded as a functional equation for
$\Lambda(u)$.

\medskip
\noindent
{\it Crossing symmetry}
\medskip

A second functional equation can be obtained from the crossing relation
$$t(u) = t(-u -\rho) \,, \eqlabel{crossing-transfer} $$
which is proved in Appendix C. This immediately leads to the desired
equation
$$\Lambda(u) = \Lambda(-u -\rho) \,. \eqlabel{crossing-lambda} $$

\medskip
\noindent
{\it Asymptotic behavior}
\medskip

The leading asymptotic
behavior of $\Lambda(u)$ for large $u$ is given by
$$\Lambda(u) \sim \kappa(u)^2
\tr_a\ M_a <\Lambda|\ T^+_a\ \hat T^+_a\ |\Lambda> = \kappa(u)^2
\tr_a\ M_a^{-1} <\Lambda|\ \hat T^+_a\ T^+_a\ |\Lambda> \quad
{\hbox{  for  }} u \rightarrow \infty \,. \eqlabel{asymptoticLambda}
$$
Here we have used Eqs. \eqref{transfer}, \eqref{transfertoo}
and \eqref{asymptoticT},
and we have assumed that the state $|\Lambda>$ is normalized to unity.
As already remarked, the matrices
$T^+_a$ and $\hat T^+_a$ are triangular matrices whose off-diagonal
elements are either raising or lowering operators of the quantum algebra
$U_q[g_0]$. We shall require $|\Lambda>$ to be a highest-weight vector
of this algebra. It follows that a trace in
\eqref{asymptoticLambda} is over a {\it diagonal} matrix involving only
the Cartan generators, which is easily evaluated.

\medskip
\noindent
{\it Periodicity and analyticity}
\medskip

{}From the periodicity and analyticity properties of $R(u)$ as a function of
$u$,
one can readily deduce the corresponding properties of $\Lambda(u)$.

\bigskip

The equations \eqref{fusion-lambda} and \eqref{crossing-lambda},
the asymptotic behavior, and the periodicity and analyticity properties
outlined above are the basic ingredients of the analytical Bethe Ansatz
method. Taking a ``dressed'' pseudovacuum eigenvalue as an Ansatz,
these properties can be used to completely determine $\Lambda(u)$.
In Section 4, we work out the simplest case, $A^{(1)}_1$. Although this model
has already been solved by other approaches, we perform this exercise
in order both to check and to illustrate the new method. In Section 5,
we work out the next-simplest case, $A^{(2)}_2$, which previously has not
been solved. This calculation already exhibits most of the complications
which are present in the general case.

\vskip 0.4truein
\noindent
{\bf \chapnum . The case $A^{(1)}_1$}
\vskip 0.2truein

The $R$ matrix associated with the spin 1/2 representation of $A^{(1)}_1$
is given by
$$R(u)= \left( \matrix{ \sh(u+\eta)                         \cr
                         &  \sh u          & e^u \sh \eta  \cr
                         & e^{-u} \sh \eta  & \sh u        \cr
                         &                  & & \sh(u+\eta) \cr} \right) \,.
\eqlabel{R(1)} $$
This form of the $R$ matrix differs from the more familiar symmetric form
(which
does not satisfy the commutativity property
\eqref{Rcheck}) by a gauge transformation${}^{\refref{jimbo},
\refref{gauge},\refref{jpa}}$.
It satisfies the crossing property \eqref{crossing} with
$$V = \left( \matrix{ & -ie^{-\eta/2} \cr
                      ie^{\eta/2}     \cr} \right) \,, \quad\quad
  M = V^t\ V = -\left( \matrix{ e^{\eta} \cr
                      & e^{-\eta}       \cr} \right) \,, \eqlabel{V/M(1)}
$$
and $\rho = \eta + i\pi$. Moreover,
$$\g(u) = \sh(-u + \eta) \,, \quad\quad\quad
     \zeta(u) = \sh(u  + \eta)\ \sh(-u + \eta) \,, \eqlabel{g/zeta(1)} $$
where $\g(u)$ is defined in \eqref{calg}; and
$$\alpha(u) = \sh(u + \eta) \,, \quad\quad\quad
   \beta(u) = \sh(2u + 2\eta) \,, \eqlabel{alpha/beta(1)} $$
where $\alpha(u)$ and $\beta(u)$ are introduced in \eqref{fusedR} and
\eqref{fusedK}, respectively.
This $R$ matrix has the further property
$$R(u + i\pi) = - R(u) \,, \eqnum $$
which implies that the transfer matrix \eqref{transfer} is a periodic function
of $u$, with period $i\pi$,
$$t(u + i\pi) = t(u) \,. \eqlabel{periodicity(1)} $$

We now determine the leading asymptotic behavior of $t(u)$ for large $u$.
It is easy to see that for $u\rightarrow \infty$,
$$\eqalignno{
  &R_{ak}(u) \sim {1\over 2} e^{u + {\eta\over 2}}
\left( \matrix{ e^{\eta S^3_k}  & p\ S^-_k \cr
                     0           & e^{-\eta S^3_k} \cr} \right) \,, \cr
  & \quad \cr
  &R_{ka}(u) \sim {1\over 2} e^{u + {\eta\over 2}}
\left( \matrix{ e^{\eta S^3_k}  & 0  \cr
p\ S^+_k  & e^{-\eta S^3_k} \cr} \right) \,,
\quad\quad k = 1\,, \cdots \,, N \,, \eqalignnum \cr} $$
where
$$ p = 2 e^{-{\eta\over 2}} \sh \eta \,, \eqnum $$
and $S^3_k$, $S^\pm_k$ are given by \eqref{spins(1)}. (Here we write the $R$
matrix \eqref{R(1)} as a $2 \times 2$ matrix in the auxiliary space,
with operator entries.) It follows that
$$T_{a}(u) \sim \kappa(u)\  T^+_a  = \left({1\over 2}\right)^N
e^{(u + {\eta\over 2}) N}
\left( \matrix{ e^{\eta S^3}  & p\ S^- \cr
                     0         & e^{-\eta S^3} \cr} \right) \,,
$$

$$\hat T_{a}(u) \sim \kappa(u)\ \hat T^+_a = \left({1\over 2}\right)^N
e^{(u + {\eta\over 2}) N}
\left( \matrix{ e^{\eta S^3}  & 0  \cr
p\ S^+  & e^{-\eta S^3} \cr} \right) \,,
\eqnum $$
where $S^3$, $S^\pm$ are given by \eqref{comult}, and $q=e^\eta$.
For later convenience, we now introduce the operator ${\cal M}$, defined by
$$ {\cal M} = {N\over 2} - S^3 \,. \eqlabel{calM(1)} $$
Recalling the expression \eqref{transfer} for $t(u)$, we conclude that its
leading asymptotic behavior for large $u$ is given by
$$t(u) \sim - \left( {1\over 2}\right)^{2N}\ e^{2 u N}\ \left\{
e^{\eta (1 + 2N - 2{\cal M})} + e^{\eta (-1 + 2 {\cal M})} + 4 e^{\eta N}
\sh^2 \eta\ S^-\ S^+ \right\} \,. \eqlabel{step} $$

As discussed in Section 2, $t(u)$ commutes with $T^+_a$ and
$\hat T^+_a$,
and therefore with $S^\pm$ and ${\cal M}$. Let $|\Lambda^{(m)}>$ be
simultaneous eigenstates of $t(u)$ and ${\cal M}$,
$$\eqalignno{
    t(u)\ |\Lambda^{(m)}> &= \Lambda^{(m)}(u)\ |\Lambda^{(m)}> \,, \cr
{\cal M}\ |\Lambda^{(m)}> &= m\ |\Lambda^{(m)}> \,. \eqalignnum \cr} $$
We choose these eigenstates to be highest weights of $U_q[su(2)]$,
$$ S^+\ |\Lambda^{(m)}> = 0 \,. \eqlabel{highestweight(1)} $$
(Within the algebraic Bethe Ansatz approach, one can
prove${}^{\refref{mpla}, \refref{devega}}$
that the Bethe states of this model are highest
weights of $U_q[su(2)]$.) It follows from \eqref{step} that the leading
asymptotic behavior of $\Lambda^{(m)}(u)$ for large $u$ is given by
$$\Lambda^{(m)}(u) \sim - \left({1\over 2}\right)^{2N}\ e^{2 u N}\
\left\{ e^{\eta (1 + 2N - 2m)} + e^{\eta (-1 + 2m)}  \right\} \,.
\eqlabel{asymptotic(1)} $$

Consider the so-called pseudovacuum state which has all $N$ spins up:
$$|\uparrow \uparrow \cdots \uparrow> = \prod_{k=1}^N \otimes |\uparrow>_k \,,
\eqlabel{pseudovacuum} $$
where
$$|\uparrow>_k = \left( \matrix{1 \cr
                                0 \cr} \right)_k \,. \eqnum $$
This state clearly satisfies the highest-weight
condition \eqref{highestweight(1)}, and has $m=0$. Let us assume that
this is an eigenstate $|\Lambda^{(0)}>$ of $t(u)$. Normalizing this state
to unity, it follows that the eigenvalue $\Lambda^{(0)}(u)$ is given by
$$\eqalignno{
\Lambda^{(0)}(u) &= <\Lambda^{(0)}| \ t(u)\ |\Lambda^{(0)}> \cr
                 &= \tr_a \ M_a <\Lambda^{(0)}| \ T_a(u)\ \hat T_a(u)\
  |\Lambda^{(0)}> \,. \eqalignlabel{matrixelement} \cr} $$
In order to evaluate this matrix element, we observe that
${}_k<\uparrow|\ R_{ak}(u)$ and $R_{ka}(u)\ |\uparrow>_k$ are lower- and
upper- triangular matrices, respectively:
$${}_k<\uparrow|\ R_{ak}(u) = {}_k<\uparrow|\
\left( \matrix{w_0 + w_3\ \sigma^3_k & 0 \cr
w_+\ \sigma^+_k & w_0 - w_3\  \sigma^3_k \cr} \right) \,, $$
$$R_{ka}(u)\ |\uparrow>_k =
\left( \matrix{w_0 + w_3\  \sigma^3_k
& w_- \ \sigma^-_k \cr
0 & w_0 - w_3\ \sigma^3_k \cr} \right)\ |\uparrow>_k
\,, \eqnum $$
where
$$
  w_0 = \sh(u+{\eta\over 2}) \ch{\eta\over 2} \,, \quad\quad\quad
  w_3 = \ch(u+{\eta\over 2}) \sh{\eta\over 2} \,, \quad\quad\quad
  w_+ = w_- = e^{-u} \sh\eta  \,, \eqnum $$
and
$\sigma^\pm_k = (\sigma^1_k \pm i \sigma^2_k)/2$.
Hence, $<\Lambda^{(0)}|\ T_{a}(u)$ and $\hat T_{a}(u)\ |\Lambda^{(0)}>$ are
also
lower- and upper- triangular matrices, which can be readily computed. We find
$$\eqalignno{
\Lambda^{(0)}(u) &= -\left\{ e^\eta \sh^{2N}(u + \eta) +
e^{-\eta} \left[ \sh^{2N} u + e^{-2u} \sum_{k=1}^N \sh^{2(N-k)}(u + \eta)\
\sh^{2(k-1)}u \right] \right\} \cr
&= -{1\over \sh(2u + \eta)} \left\{ \sh(2u + 2\eta)\ \sh^{2N}(u + \eta)
+ \sh2u\ \sh^{2N} u \right\} \,. \eqalignnum \cr} $$

Assuming that a general eigenvalue $\Lambda^{(m)}(u)$ has the form of a
``dressed'' pseudovacuum eigenvalue, we are led to the following analytical
Ansatz,
$$\Lambda^{(m)}(u) =-{1\over \sh(2u + \eta)} \left\{
A^{(m)}(u)\ \sh(2u + 2\eta)\ \sh^{2N}(u + \eta)
+ B^{(m)}(u)\ \sh2u\ \sh^{2N}u \right\} \,. \eqlabel{ansatz(1)} $$
The functions $A^{(m)}(u)$ and $B^{(m)}(u)$ are to determined with the
help of the functional equations of Section 3, as well as the asymptotic
behavior for large $u$
$$A^{(m)}(u) \rightarrow e^{-2m\eta} \,, \quad\quad\quad
  B^{(m)}(u) \rightarrow e^{2m\eta} \eqlabel{ABasymptotic} $$
which follows from \eqref{asymptotic(1)}.

Let us substitute the above Ansatz into the fusion equation
\eqref{fusion-lambda}, making use of \eqref{g/zeta(1)} and
\eqref{alpha/beta(1)}.
The fact that the resulting equation is identically satisfied for
$A^{(0)} = B^{(0)} = 1$ confirms our earlier assumption that the
pseudovacuum state is an eigenstate of $t(u)$. Moreover, we obtain the
result
$$A^{(m)}(u + \rho)\ B^{(m)}(u) = 1 \,. \eqlabel{ABfusion} $$

Similarly, substituting the Ansatz into the crossing relation
\eqref{crossing-lambda}, we obtain
$$ A^{(m)}(u) = B^{(m)}(-u - \rho) \,. \eqlabel{ABcrossing} $$
Combining these two results, we deduce that
$$A^{(m)}(u)\ A^{(m)}(-u) = 1 \,. \eqlabel{AA(1)} $$

Introduce the variable $\lambda = e^u$. The fact that the $u$-dependence
of the $R$ matrix \eqref{R(1)} involves only $\lambda^{\pm 1}$ and
$\lambda^0$ implies that the transfer matrix \eqref{transfer}
can be expanded as a finite power series in $\lambda$,
$$t(u) = \sum_{i=-N}^N t_i\ \lambda^{2i} \,, \eqnum $$
where $\{ t_i \}$ are operators. Because of the periodicity property
\eqref{periodicity(1)}, only even powers of $\lambda$ appear in this sum.
The commutativity property \eqref{commutativity} implies
$[t_i \,, t_j ] = 0$ for all $i \,, j$. Setting
$$t_i\ |\Lambda^{(m)}> = c_i^{(m)}\ |\Lambda^{(m)}> \,, \quad\quad
i = -N \,, \cdots \,, N \,, \eqnum $$
we conclude that
$$ \Lambda^{(m)}(u) = \sum_{i=-N}^N c_i^{(m)}\ \lambda^{2i} \,. \eqnum $$
That is, $\Lambda^{(m)}(u)$ is a finite power series in $\lambda^2$; in
particular, it has poles only at $\lambda = 0$ and $\lambda = \infty$.
(This argument is a direct generalization of the one given by
Reshetikhin${}^{\refref{reshetikhin}}$ for the corresponding closed chain.)

In view of the Ansatz \eqref{ansatz(1)}, we see that $A^{(m)}$ and $B^{(m)}$
must be rational functions of $\lambda^2$; they must have common poles, and the
residues of $\Lambda^{(m)}(u)$ at these poles must vanish. Setting
$$A^{(m)}(u) = a \prod_{j=1}^m {\lambda^2 - \alpha_j \over \lambda^2 - \beta_j}
{\lambda^2 - \gamma_j \over \lambda^2 - \delta_j} \,, \eqnum $$
we find that the asymptotic behavior \eqref{ABasymptotic} implies
$a=e^{-2m\eta}$, and that the condition \eqref{AA(1)} is consistent with
$$\beta_j = {1\over \alpha_j} \,, \quad\quad \gamma_j = \pm {e^{2\eta}\over
\alpha_j} \,, \quad\quad \delta_j = \pm e^{-2\eta}\alpha_j \,, \quad\quad
j = 1\,, \cdots \,, m \,. \eqnum $$

Imposing the condition that $B^{(m)}(u) = A^{(m)}(-u -\rho)$ has the same
poles as $A^{(m)}(u)$, we arrive at the expressions
$$\eqalignno{
A^{(m)}(u) &= e^{-2m\eta} \prod_{j=1}^m {\lambda^2 - \alpha_j \over
\lambda^2 -1/ \alpha_j}
{\lambda^2 - e^{2\eta}/\alpha_j \over \lambda^2 - e^{-2\eta}\alpha_j} \,, \cr
B^{(m)}(u) &= e^{2m\eta} \prod_{j=1}^m {\lambda^2 - e^{-2\eta}/\alpha_j \over
\lambda^2 -1/ \alpha_j}
{\lambda^2 - e^{-4\eta}\alpha_j \over \lambda^2 - e^{-2\eta}\alpha_j} \,.
\eqalignnum \cr} $$
The condition that the residue of $\Lambda^{(m)}(u)$ at
$\lambda^2 = 1/\alpha_k \equiv e^{2u_k - \eta}$ vanishes leads to the
Bethe Ansatz (BA) equations
$$\left[ {\sh(u_k + {\eta\over 2}) \over \sh(u_k - {\eta\over 2})} \right]^{2N}
= \prod_{j \ne k} {\sh(u_k - u_j + \eta) \over \sh(u_k - u_j - \eta)}
{\sh(u_k + u_j + \eta) \over \sh(u_k + u_j - \eta)} \,, \quad\quad
k = 1\,, \cdots \,, m \,. \eqlabel{BA(1)} $$
(We have made the shift by $\eta$ in order to bring these equations to a more
symmetric form.)
Moreover, we conclude that the eigenvalues $\Lambda^{(m)}(u)$ are given by
$$\eqalignno{
\Lambda^{(m)}(u) =-{1\over \sh(2u + \eta)} & \biggl\{
\sh(2u + 2\eta)\ \sh^{2N}(u + \eta)\ \prod_{j=1}^m
{\sh(u - u_j - {\eta\over 2}) \over \sh(u - u_j + {\eta\over 2})}
{\sh(u + u_j - {\eta\over 2}) \over \sh(u + u_j + {\eta\over 2})} \cr
&+  \sh2u\ \sh^{2N}u \prod_{j=1}^m
{\sh(u - u_j + {3\eta\over 2}) \over \sh(u - u_j + {\eta\over 2})}
{\sh(u + u_j + {3\eta\over 2}) \over \sh(u + u_j + {\eta\over 2})}
\biggr\} \,, \eqalignlabel{spectrum(1)} \cr} $$
where $\left\{u_1 \,, \cdots u_m \right\}$ are solutions of the BA equations
\eqref{BA(1)}.
While we have not fully justified some of the assumptions that we have used,
the fact that these results coincide with those obtained in
Refs. \refref{alcaraz} and \refref{sklyanin} (using the coordinate BA and
the algebraic BA, respectively) suggests that this method is sound.

\vskip 0.4truein
\noindent
{\bf \chapnum . The case $A^{(2)}_2$}
\vskip 0.2truein

The $R$ matrix associated with the fundamental representation of $A^{(2)}_2$
is given by${}^{\refref{izergin},\refref{kulish/sklyanin(3)}}$
$$R(u) = \left( \matrix{c                             \cr
                        & b & & e                     \cr
                        & & d & & g & & f             \cr
                        & \bar e & & b                \cr
                        & & \bar g & & a & & g        \cr
                        & & & & & b & & e             \cr
                        & & \bar f & & \bar g &  & d  \cr
                        & & & & & \bar e & & b        \cr
                        & & & & & & & & c             \cr} \right) \,,
\eqlabel{R(2)}  $$
where
$$\eqalignno{
a & = \sh (u - 3\eta) - \sh 5\eta + \sh 3\eta + \sh \eta \,, \quad\quad
b =  \sh (u - 3\eta) + \sh 3\eta  \,,  \cr
c & = \sh (u - 5\eta) + \sh \eta  \,, \quad\quad\quad\quad\quad\quad\quad\quad
d =  \sh (u - \eta) + \sh \eta \,,  \cr
e & = -2 e^{-{u\over 2}} \sh 2\eta \ \ch ({u\over 2} - 3\eta)  \,,
\quad\quad\quad\quad\quad
\bar e = -2 e^{{u\over 2}} \sh 2\eta \ \ch ({u\over 2} - 3\eta)  \,, \cr
f &= -2 e^{-u + 2\eta} \sh \eta \ \sh 2\eta -  e^{-\eta} \sh 4\eta \,,
\quad\quad
\bar f = 2 e^{u - 2\eta} \sh \eta \ \sh 2\eta -  e^{\eta} \sh 4\eta \,, \cr
g & = 2 e^{-{u\over 2} + 2\eta} \sh {u\over 2} \ \sh 2\eta \,,
\quad\quad\quad\quad\quad\quad\quad
\bar g = - 2 e^{{u\over 2} - 2\eta} \sh {u\over 2} \ \sh 2\eta \,.
\eqalignlabel{abcdefg} \cr} $$
This form of the $R$ matrix differs from the one given in Ref.
\refref{kulish/sklyanin(3)} by a similarity transformation, as well as by an
overall multiplicative factor. The corresponding crossing matrices are given
by
$$ V = \left( \matrix{&   &  -e^{-\eta}   \cr
                      &   1               \cr
                      -e^{\eta}           \cr} \right) \,, \quad\quad
M = V^t\ V = \left( \matrix{e^{2\eta}   \cr
                      &   1             \cr
                      &   &  e^{-2\eta} \cr } \right) \,, \eqlabel{V/M(2)} $$
with $\rho = -6\eta - i \pi$. Furthermore,
$$\eqalignno{
  \g(u) &= 2\ch({u\over 2} + 3\eta)\ \sh({u\over 2} - 2\eta) \,, \cr
    \zeta (u) &= -4\ch({u\over 2} + 3\eta)\ \ch({u\over 2} - 3\eta)\
\sh({u\over 2} + 2\eta)\ \sh({u\over 2} - 2\eta) \,, \cr
\alpha(u) &= \ch({u\over 2} - 3\eta) \,, \quad\quad\quad
 \beta(u) = \sh(u - 6\eta) \,. \eqalignlabel{g/zeta(2)} \cr} $$
Also,
$$R(u + 2i\pi) = R(u) \,, \eqnum $$
which implies that
$$t(u + 2i\pi) = t(u) \,. \eqlabel{periodicity(2)} $$

We observe that for $u \rightarrow \infty$,
$$T_{a}(u) \sim \left( {1\over 2} \right)^N
e^{(u - 3\eta) N}
\left( \matrix{
e^{-2\eta S^3}                                         & 0 & 0 \cr
p_1\ S^+ e^{-\eta S^3}  & 1 & 0 \cr
p_2\ (S^+)^2  & p_1
e^{\eta S^3} S^+ & e^{2\eta S^3} \cr} \right) \,,
$$
$$\hat T_{a}(u) \sim \left({1\over 2}\right)^N
e^{(u - 3\eta) N}
\left( \matrix{
e^{-2\eta S^3} & p_1\
e^{-\eta S^3} S^-   & p_2 (S^-)^2  \cr
0 & 1 & p_1\ S^- e^{\eta S^3}  \cr
0 & 0 & e^{2\eta S^3} \cr} \right) \,,  \eqnum
$$
where
$$p_1 = -{2\sh 2\eta\over{\sqrt{2\ch\eta}}} \,, \quad\quad\quad
  p_2 = 4\sh^2 \eta\ e^\eta  \,, \eqnum $$
and $S^3$, $S^\pm$ are given by \eqref{comult}, with $q=e^\eta$,
and
$$S^3_k = \left( \matrix{1       \cr
                         & 0     \cr
                         &  &-1  \cr} \right) \,, \quad\quad
S^+_k = {\sqrt{2\ch \eta}} \left( \matrix{\ &1  &\  \cr
                                          \ &\  &1  \cr
                                          \ &\  &\  \cr} \right) \,, \quad\quad
S^-_k = {\sqrt{2\ch \eta}} \left( \matrix{\ &\  &\  \cr
                                          1 &\  &\  \cr
                                          \ &1  &\  \cr} \right) \,,
\eqlabel{spins(2)} $$
which we recognize as the spin-1 representation of $U_q[su(2)]$.
The fact \eqref{commute} that $T^+_a$ and $\hat T^+_a$ commute with
$t(u)$ provides an explicit proof of the $U_q[su(2)]$-invariance of the
transfer matrix.

As in the $A^{(1)}_1$ case, there exist simultaneous eigenstates of $t(u)$
and ${\cal M}$,
$$\eqalignno{
    t(u)\ |\Lambda^{(m)}> &= \Lambda^{(m)}(u)\ |\Lambda^{(m)}> \,, \cr
{\cal M}\ |\Lambda^{(m)}> &= m\ |\Lambda^{(m)}> \,, \eqalignnum \cr} $$
where ${\cal M}$ is now defined by
$$ {\cal M} = N - S^3 \,. \eqlabel{calM(2)} $$
Again, we choose these to be highest weights of $U_q[su(2)]$,
$$ S^+\ |\Lambda^{(m)}> = 0 \,. \eqlabel{highestweight(2)} $$
Evaluating the second matrix element in \eq\eqref{asymptoticLambda},
we find that the leading asymptotic
behavior of $\Lambda^{(m)}(u)$ for large $u$ is given by
$$\Lambda^{(m)}(u) \rightarrow \left( {1\over 2} \right)^{2N}\ e^{2 u N}\
\left\{ e^{2\eta (1 - N - 2m)} + e^{-6\eta N} + e^{2\eta (-1 - 5N + 2m)}
\right\} \,.
\eqlabel{asymptotic(2)} $$

We assume that the pseudovacuum state
$|\uparrow \uparrow \cdots \uparrow> = |\Lambda^{(0)}>$, which is
given by \eqref{pseudovacuum} with
$$|\uparrow>_k = \left( \matrix{1 \cr
                                0 \cr
                                0 \cr} \right)_k \,, \eqnum $$
is an eigenstate of $t(u)$; and we proceed to calculate the corresponding
eigenvalue $\Lambda^{(0)}(u)$. We observe that
$${}_k<\uparrow|\ R_{ak}(u) = {}_k<\uparrow|\
\left( \matrix{A_k & 0   & 0   \cr
               E_k & B_k & 0   \cr
               F_k & G_k & C_k \cr} \right) \,, $$
$$R_{ka}(u)\ |\uparrow>_k =
\left( \matrix{A_k & \bar E_k & \bar F_k \cr
               0   &      B_k & \bar G_k \cr
               0   &        0 &      C_k \cr} \right)\ |\uparrow>_k
\,, \eqnum $$
where
$$\eqalignno{
 A_k &= \left(\matrix{c     \cr
                      & b   \cr
                      & & d \cr} \right) \,, \quad\quad
  B_k = \left(\matrix{b     \cr
                      & a   \cr
                      & & b \cr} \right) \,, \quad\quad
  C_k = \left(\matrix{d     \cr
                      & b   \cr
                      & & c \cr} \right) \,, \cr
 E_k &= \left(\matrix{\ & \bar e &\       \cr
                      \ &\       & \bar g \cr
                      \ &\       &\       \cr} \right) \,, \quad\quad
  F_k = \left(\matrix{\ &\       & \bar f \cr
                      \ &\       &\       \cr
                      \ &\       &\       \cr} \right) \,, \quad\quad
  G_k = \left(\matrix{\ & \bar g &\       \cr
                      \ &\       & \bar e \cr
                      \ &\       &\       \cr} \right) \,, \eqalignnum \cr
 \bar E_k &= \left(\matrix{\       &\        &\    \cr
                           \bar e  &\        &\    \cr
                           \       & \bar g  &\    \cr} \right) \,, \quad\quad
  \bar F_k = \left(\matrix{\       &\        &\    \cr
                           \       &\        &\    \cr
                           \bar f  &\        &\    \cr} \right) \,, \quad\quad
  \bar G_k = \left(\matrix{\       &\        &\    \cr
                           \bar g  &\        &\    \cr
                           \       & \bar e  &\    \cr} \right) \,,
\quad k = 1\,, \cdots \,, N \,.  \cr}$$
Therefore,
$$<\Lambda^{(0)}|\ T_{a}(u) = <\Lambda^{(0)}|\
\left( \matrix{{\cal A}_N &        0   & 0          \cr
               {\cal E}_N & {\cal B}_N & 0          \cr
               {\cal F}_N & {\cal G}_N & {\cal C}_N \cr} \right) \,,
\quad\quad
\hat T_{a}(u)\ |\Lambda^{(0)}> =
\left( \matrix{{\cal A}_N & \bar {\cal E}_N & \bar {\cal F}_N \cr
               0   &      {\cal B}_N & \bar {\cal G}_N \cr
               0   &        0 &  {\cal C}_N \cr} \right)\ |\Lambda^{(0)}> \,,
\eqnum $$
where
$${\cal A}_N = A_N\ A_{N-1} \cdots A_1 \,, \quad\quad
  {\cal B}_N = B_N\ B_{N-1} \cdots B_1 \,, \quad\quad
  {\cal C}_N = C_N\ C_{N-1} \cdots C_1 \,, \eqnum $$
$$\eqalignno{
{\cal E}_N &= \sum_{k=1}^N B_N \cdots B_{k+1}\ E_k\ A_{k-1} \cdots A_1
\,, \quad\quad {\cal E}_0 = 0  \,, \eqalignlabel{ee} \cr
{\cal G}_N &= \sum_{k=1}^N C_N \cdots C_{k+1}\ G_k\ B_{k-1} \cdots B_1
\,, \quad\quad {\cal G}_0 = 0  \,, \eqalignlabel{gg} \cr
{\cal F}_N &= \sum_{k=1}^N C_N \cdots C_{k+1}\ \left( F_k\ A_{k-1} \cdots A_1
+ G_k\ {\cal E}_{k-1} \right) \,,  \eqalignlabel{ff} \cr} $$
and $\bar {\cal E}_N$, $\bar {\cal G}_N$, $\bar {\cal F}_N$ are given by
expressions \eqref{ee}, \eqref{gg}, \eqref{ff} with $E_k$, $G_k$, $F_k$ and
${\cal E}_{k-1}$ replaced by $\bar E_k$, $\bar G_k$, $\bar F_k$ and
$\bar {\cal E}_{k-1}$, respectively.
The corresponding matrix element \eqref{matrixelement} is therefore
$$\Lambda^{(0)}(u) = < \Lambda^{(0)}| \left\{ e^{2\eta} {\cal A}_N^2 +
{\cal E}_N \bar {\cal E}_N  + {\cal B}_N^2
+ e^{-2\eta} \left(  {\cal F}_N \bar {\cal F}_N
+ {\cal G}_N \bar {\cal G}_N + {\cal C}_N^2 \right)
\right\} |\Lambda^{(0)}> \,. \eqnum $$
After a lengthy calculation, we conclude that
$$\eqalignno{
\Lambda^{(0)}(u) &= c^{2N} \left\{ e^{2\eta} - {\bar e^2\over b^2 - c^2}
+ e^{-2\eta}\left[ {\bar f^2\over c^2 - d^2}
- {\bar e^2 \bar g^2\over (b^2 - c^2)(c^2 - d^2)} \right] \right\} \cr
&+ b^{2N} \left\{ 1 + {\bar e^2\over b^2 - c^2}
+ e^{-2\eta}\left[ {\bar g^2\over b^2 - d^2}
+ {\bar e^2 \bar g^2\over (b^2 - c^2)(b^2 - d^2)} \right] \right\} \cr
&+ d^{2N}\ e^{-2\eta} \left\{ 1 - {\bar g^2\over b^2 - d^2}
- {\bar f^2\over c^2 - d^2}
+ {\bar e^2 \bar g^2\over (b^2 - d^2)(c^2 - d^2)} \right\} \cr
&= c^{2N} {\sh(u - 6\eta) \ch(u - \eta) \over\sh(u - 2\eta) \ch(u -3\eta)}
+  b^{2N} {\sh u \sh(u - 6\eta) \over\sh(u - 2\eta) \sh(u -4\eta)} \cr
&\quad\quad\quad\quad
+ d^{2N} {\sh u \ch(u - 5\eta) \over\sh(u - 4\eta) \ch(u -3\eta)}
\,, \eqalignnum \cr}$$
where $b$, $c$, $d$, etc. are given in \eq\eqref{abcdefg}.

The assumption that a general eigenvalue $\Lambda^{(m)}(u)$ has the form
of a ``dressed'' pseudovacuum eigenvalue leads to the analytical Ansatz
$$\eqalignno{
\Lambda^{(m)}(u)
&= A^{(m)}(u)\ c^{2N}
{\sh(u - 6\eta) \ch(u - \eta) \over\sh(u - 2\eta) \ch(u -3\eta)}
+  B^{(m)}(u)\ b^{2N}
{\sh u \sh(u - 6\eta) \over\sh(u - 2\eta) \sh(u -4\eta)} \cr
&\quad\quad\quad\quad
+ C^{(m)}(u)\ d^{2N}
{\sh u \ch(u - 5\eta) \over\sh(u - 4\eta) \ch(u -3\eta)}
\,, \eqalignlabel{ansatz(2)} \cr}$$
where the functions $A^{(m)}(u)$, $B^{(m)}(u)$ and $C^{(m)}(u)$ are to be
determined. From \eqref{asymptotic(2)}, we obtain the asymptotic behavior
for large $u$
$$A^{(m)}(u) \rightarrow e^{4\eta m} \,, \quad\quad
  B^{(m)}(u) \rightarrow 1           \,, \quad\quad
  C^{(m)}(u) \rightarrow e^{-4\eta m} \,. \eqlabel{ABCasymptotic}
$$
The condition that the residue of $\Lambda^{(m)}(u)$ at $u=2\eta$ vanishes
implies
$$A^{(m)}(2\eta) = B^{(m)}(2\eta) \,. \eqlabel{vanish} $$

We now substitute this Ansatz into the fusion equation \eqref{fusion-lambda},
making use of \eqref{g/zeta(2)}. The resulting equation is identically
satisfied for $A^{(0)} = B^{(0)} = C^{(0)} = 1$, thereby confirming that the
pseudovacuum state is an eigenstate of $t(u)$. Moreover, we obtain
$$A^{(m)}(u + \rho)\ C^{(m)}(u) = 1 \,. \eqlabel{ACfusion} $$
The crossing relation \eqref{crossing-lambda} implies
$$ C^{(m)}(u) = A^{(m)}(-u - \rho) \,, \eqlabel{ACcrossing} $$
$$ B^{(m)}(u) = B^{(m)}(-u - \rho) \,. \eqlabel{BBcrossing} $$
{}From \eqref{ACfusion} and \eqref{ACcrossing}, we see that
$$A^{(m)}(u)\ A^{(m)}(-u) = 1 \,. \eqlabel{AA(2)} $$

As in the $A^{(1)}_1$ case, the $R$ matrix \eqref{R(2)} involves only
$\lambda^{\pm 1}$ and $\lambda^0$, where $\lambda = e^u$. It follows that
$\Lambda^{(m)}(u)$ is a finite power series in $\lambda$, and has poles
only at $\lambda = 0$ and $\lambda = \infty$. Thus, $A^{(m)}$, $B^{(m)}$
and $C^{(m)}$ must be rational functions of $\lambda$, and the
residues of $\Lambda^{(m)}(u)$ must vanish. We assume
$$A^{(m)}(u) = a \prod_{j=1}^m {\lambda - \alpha_j \over \lambda - \beta_j}
{\lambda - \gamma_j \over \lambda - \delta_j} \,. \eqnum $$
The asymptotic behavior \eqref{ABCasymptotic} implies
$a=e^{4 m \eta}$. The condition \eqref{AA(2)} is consistent with
$$\beta_j = {1\over \alpha_j} \,, \quad\quad \gamma_j = \pm {e^{-4\eta}\over
\alpha_j} \,, \quad\quad \delta_j = \pm e^{4\eta}\alpha_j \,, \quad\quad
j = 1\,, \cdots \,, m \,. \eqnum $$

Arranging for $B^{(m)}(u)$ to have the same poles as $A^{(m)}(u)$ and
$C^{(m)}(u)$, and imposing the condition \eqref{vanish}, we obtain
$$\eqalignno{
A^{(m)}(u) &= e^{4 m\eta} \prod_{j=1}^m {\lambda - \alpha_j \over
\lambda -1/ \alpha_j}
{\lambda - e^{-4\eta}/\alpha_j \over \lambda - e^{4\eta}\alpha_j} \,, \cr
B^{(m)}(u) &= \prod_{j=1}^m {\lambda + e^{2\eta} \alpha_j \over
\lambda -1/ \alpha_j}
{\lambda - e^{4\eta}/\alpha_j \over \lambda - e^{4\eta}\alpha_j}
{\lambda + e^{-2\eta}/\alpha_j \over \lambda + e^{6\eta} \alpha_j}
{\lambda - e^{8\eta}\alpha_j \over \lambda + e^{2\eta}/\alpha_j} \,, \cr
C^{(m)}(u) &= e^{-4m\eta} \prod_{j=1}^m {\lambda + e^{6\eta}/\alpha_j \over
\lambda + e^{6\eta} \alpha_j}
{\lambda + e^{10\eta}\alpha_j \over \lambda + e^{2\eta}/\alpha_j} \,.
\eqalignnum \cr} $$
The condition that the residue of $\Lambda^{(m)}(u)$ at
$\lambda = 1/\alpha_k \equiv e^{u_k + 2\eta}$ vanishes leads to the
BA equations
$$\eqalignno{
\left[ {\sh({u_k\over 2} - \eta) \over \sh({u_k\over 2} + \eta)} \right]^{2N}
& = \prod_{j \ne k} \biggl\{
{\sh \left[{1\over 2}(u_k + u_j) - 2\eta \right] \over
 \sh \left[{1\over 2}(u_k + u_j) + 2\eta \right]}
{\ch \left[{1\over 2}(u_k + u_j) + \eta \right] \over
 \ch \left[{1\over 2}(u_k + u_j) - \eta \right]} \cr
& \cdot
{\sh \left[{1\over 2}(u_k - u_j) - 2\eta \right] \over
 \sh \left[{1\over 2}(u_k - u_j) + 2\eta \right]}
{\ch \left[{1\over 2}(u_k - u_j) + \eta \right] \over
 \ch \left[{1\over 2}(u_k - u_j) - \eta \right]} \biggr\}
\,, \quad k = 1\,, \cdots \,, m \,. \eqalignlabel{BA(2)} \cr} $$
Furthermore, the eigenvalues $\Lambda^{(m)}(u)$ are given by
\eqref{ansatz(2)}, with
$$\eqalignno{
A^{(m)}(u) &= \prod_{j=1}^m
{\sh \left[{1\over 2}(u + u_j) + \eta \right] \over
 \sh \left[{1\over 2}(u + u_j) - \eta \right]}
{\sh \left[{1\over 2}(u - u_j) + \eta \right] \over
 \sh \left[{1\over 2}(u - u_j) - \eta \right]}  \,, \cr
B^{(m)}(u) &= \prod_{j=1}^m
{\sh \left[{1\over 2}(u + u_j) - 3\eta \right] \over
 \sh \left[{1\over 2}(u + u_j) - \eta  \right]}
{\ch \left[{1\over 2}(u + u_j) \right] \over
 \ch \left[{1\over 2}(u + u_j) - 2\eta \right]}
{\sh \left[{1\over 2}(u - u_j) - 3\eta \right] \over
 \sh \left[{1\over 2}(u - u_j) - \eta  \right]}
{\ch \left[{1\over 2}(u - u_j) \right] \over
 \ch \left[{1\over 2}(u - u_j) - 2\eta \right]}  \,, \cr
C^{(m)}(u) &= \prod_{j=1}^m
{\ch \left[{1\over 2}(u + u_j) - 4\eta \right] \over
 \ch \left[{1\over 2}(u + u_j) - 2\eta \right]}
{\ch \left[{1\over 2}(u - u_j) - 4\eta \right] \over
 \ch \left[{1\over 2}(u - u_j) - 2\eta \right]}  \,.
\eqalignlabel{spectrum(2)} \cr} $$

\vskip 0.4truein
\noindent
{\bf \chapnum . Discussion}
\vskip 0.2truein

We have seen that the analytical BA approach can be used to determine the
spectrum of the transfer matrices of the quantum-algebra-invariant models
\eqref{hamiltonian}. Although we have worked out in detail only the cases
$A^{(1)}_1$ and $A^{(2)}_2$, the other cases listed in Section 2 can
evidently be treated in a similar way.

The quantum-algebra invariance of these models plays an important role
in this approach. Indeed, in order to determine the asymptotic behavior
of the eigenvalues $\Lambda^{(m)}(u)$ for large $u$, one uses the fact
that the corresponding eigenstates are highest weights of the quantum
algebra. Thus, open chains which do not have quantum-algebra symmetry are
presumably more difficult to solve. Indeed, the
algebraic${}^{\refref{sklyanin}}$ BA approach can be used regardless of
the presence of such a symmetry; but, this method has been applied only
to the case $A^{(1)}_1$.

It is instructive to compare our open-chain results with those for the
corresponding closed chains with periodic boundary conditions, which
have the transfer matrix
$$t(u)^{closed} = \tr_a T_a(u) \,. \eqlabel{closedtransfer} $$
For the closed spin 1/2 $A^{(1)}_1$ chain, the BA equations
are${}^{\refref{closedXXZ}}$
$$\left[ {\sh(u_k + {\eta\over 2}) \over \sh(u_k - {\eta\over 2})} \right]^{N}
= \prod_{j \ne k} {\sh(u_k - u_j + \eta) \over \sh(u_k - u_j - \eta)}
\,, \quad\quad k = 1\,, \cdots \,, m \,. \eqnum $$
Comparing with the BA equations \eqref{BA(1)} of the $U_q[su(2)]$-invariant
open chain, we see that the latter are ``doubled'' with respect to the
former. The same relationship exists in the $A^{(2)}_2$ case between the
closed-chain BA equations${}^{\refref{reshetikhin}}$ and the BA equations
\eqref{BA(2)} of the $U_q[su(2)]$-invariant open chain.
Moreover, the eigenvalues \eqref{spectrum(1)} and
\eqref{ansatz(2)},\eqref{spectrum(2)} for the
quantum-algebra-invariant chains are also doubled with respect to the
eigenvalues for the corresponding closed chains, up to certain prefactors.
(See Refs. \refref{closedXXZ} and \refref{reshetikhin}, respectively.)
In the calculation of the pseudovacuum eigenvalue, these
prefactors originate from off-diagonal elements of the monodromy matrices
$T_a(u)$ and $\hat T_a(u)$.

Evidently, this doubling phenomenon is related to the fact that the
open-chain transfer matrix \eqref{transfer} involves both $T_a(u)$ and
$\hat T_a(u)$, while the closed-chain transfer matrix \eqref{closedtransfer}
involves only $T_a(u)$.
Within the analytical BA approach, the signal of this doubling phenomenon
is the crossing relation $t(u) = t(-u - \rho)$. Since our proof of this
relation is quite general, we expect that the doubling phenomenon should
hold for all the cases enumerated in Section 2.

With this paper, we conclude our program${}^{\refref{spin1},
\refref{npb},\refref{jpa},\refref{ijmpa},\refref{mpla}}$ of constructing and
solving quantum-algebra-invariant chains with spins in higher-dimensional
representations and with larger algebras. Although we have examined in detail
only a small number of cases, we believe that the basic framework for
studying at least a large class of other cases is now in place.

An urgent problem now is to solve the BA equations for $N \rightarrow \infty$
and to determine the thermodynamic properties of these various models. This
is most interesting and subtle in the critical regime $|q|=1$, since in
this case the Hamiltonian is typically not Hermitian, and presumably
truncations on the space of states are required. We hope to address such
questions in future publications.

\bigskip

It is a pleasure to acknowledge the hospitality extended to us at the
Aspen Center for Physics and at CERN, where part of this work was performed.
We also thank M. Monti for his assistance in performing a numerical check on
one of our results.
This work was supported in part by the National Science
Foundation under Grant No. PHY-90 07517.

\vfill\eject

\chapno=-1

\noindent
{\bf Appendix \chapnum .}
\vskip 0.2truein

Here we prove the useful identity
$$\tr_a M_a\ T_a(u)\ \hat T_a(u) =
 \tr_a M_a^{-1}\ \hat T_a(u)\ T_a(u) \,. \eqlabel{useful} $$
Indeed,
$$\eqalignno{
\tr_a M_a\ T_a(u)\ \hat T_a(u) &= \tr_{ab} {\cal P}_{ab}\ M_b\ T_b(u)\
\hat T_a(u) \cr
&= \tr_{ab} \left[ {\cal P}_{ab}^{t_a}\ M_b\ T_b(u) \right]^{t_a}\
\hat T_a(u) \cr
&= \tr_{ab} {\cal P}_{ab}^{t_a}\ M_b\ T_b(u)\ \hat T_a(u)^{t_a} \cr
&= {1\over \zeta(2u + \rho)} \tr_{ab}
{\cal P}_{ab}^{t_a}\ M_b\ T_b(u)\ \hat T_a(u)^{t_a}\ R_{ba}(2u)^{t_a}\
M_a^{-1}\ R_{ab}(-2u -2\rho)^{t_a}\ M_a  \cr
&= \cdots \eqalignnum \cr} $$
where we have used the identity
$$R_{ba}(2u)^{t_a}\ M_a^{-1}\ R_{ab}(-2u -2\rho)^{t_a}\ M_a = \zeta(2u + \rho)
\eqnum $$
which follows from the unitarity and crossing properties of $R$. Noting that
$$T_b(u)\ \hat T_a(u)^{t_a}\ R_{ba}(2u)^{t_a} =
R_{ba}(2u)^{t_a}\ \hat T_a(u)^{t_a}\ T_b(u) \,, \eqnum $$
we obtain
$$\eqalignno{
\cdots &=  {1\over \zeta(2u + \rho)} \tr_{ab}
R_{ab}(-2u -2\rho)^{t_a}\ M_a\
{\cal P}_{ab}^{t_b}\ M_b\ R_{ba}(2u)^{t_a}\ \hat T_a(u)^{t_a}\ T_b(u)\
M_a^{-1}  \cr
&= \cdots  \eqalignnum \cr} $$
Making use of the degeneration of the commutativity property
\eqref{Rcheck} at $u = -\rho$ (see
Ref.\refref{npb}) and crossing symmetry, one can prove the identity
$${\cal P}_{ab}^{t_b}\ M_b\ R_{ba}(2u)^{t_a} =
R_{ba}(2u)^{t_a}\ M_a^{-1}\ {\cal P}_{ab}^{t_b} \,. \eqnum $$
Thus,
$$\eqalignno{
\cdots &=  {1\over \zeta(2u + \rho)} \tr_{ab}
R_{ab}(-2u -2\rho)^{t_a}\ M_a\
R_{ba}(2u)^{t_a}\ M_a^{-1}\ {\cal P}_{ab}^{t_b}\ \hat T_a(u)^{t_a}\ T_b(u)\
M_a^{-1}  \cr
&= \tr_{ab} {\cal P}_{ab}^{t_b}\ \hat T_a(u)^{t_a}\ T_b(u)\
M_a^{-1}  \cr
&= \tr_{ab} \hat T_a(u)^{t_a}\ M_a^{-1}\ T_b(u)\ {\cal P}_{ab}^{t_a} \cr
&= \tr_{ab} \left[ M_a^{-1}\ \hat T_a(u) \right]^{t_a}\  T_b(u)\
{\cal P}_{ab}^{t_a} \cr
&= \tr_{ab} M_a^{-1}\ \hat T_a(u)\ \left[ T_b(u)\ {\cal P}_{ab}^{t_a}
\right]^{t_a} \cr
&= \tr_{ab} M_a^{-1}\ \hat T_a(u)\ T_b(u)\ {\cal P}_{ab} \cr
&= \tr_{a} M_a^{-1}\ \hat T_a(u)\ T_a(u) \,.  \eqalignnum \cr} $$
This concludes the proof of the identity \eqref{useful}.

\vfill\eject

\noindent
{\bf Appendix \chapnum .}
\vskip 0.2truein

Here we prove that the transfer matrix \eqref{transfer} is symmetric,
$$t(u)^t = t(u) \,. \eqlabel{symmetric} $$
Indeed,
$$\eqalignno{
t(u)^t &= t(u)^{t_1 \cdots t_N} \cr
       &= \tr_a \left\{ M_a\ T_a(u)\ \hat T_a(u) \right\}^{t_a t_1 \cdots t_N}
\cr
       &= \tr_a  \hat T_a(u)^{t_a t_1 \cdots t_N}\
T_a(u)^{t_a t_1 \cdots t_N}\ M_a^{t_a}   \cr
       &= \tr_a  M_a\ \hat T_a(u)^{t_a t_1 \cdots t_N}\
T_a(u)^{t_a t_1 \cdots t_N}  \,. \eqalignlabel{step1} \cr} $$
But
$$\eqalignno{
T_a(u)^{t_a t_1 \cdots t_N} &= \left\{R_{aN}(u)\ \cdots R_{a1}(u)
\right\}^{t_a t_1 \cdots t_N} \cr
&= R_{a1}(u)^{t_a t_1}\ \cdots R_{aN}(u)^{t_a t_N} \cr
&= R_{1a}(u)\ \cdots R_{Na}(u) = \hat T_a(u) \,, \eqalignlabel{step2} \cr} $$
and similarly,
$$\hat T_a(u)^{t_a t_1 \cdots t_N} = T_a(u) \,. \eqlabel{step3} $$
Substituting \eqref{step2} and \eqref{step3} into \eqref{step1},
we conclude that
$$t(u)^t = \tr_a M_a\ T_a(u)\ \hat T_a(u)  = t(u) \,. \eqnum $$

If the $R$ matrix is real
$$R(u)^* = R(u) \,, \eqnum $$
then the transfer matrix $t(u)$ is real, in which case it is also Hermitian,
$$t(u)^\dagger = t(u) \,. \eqnum $$

\vfill\eject

\noindent
{\bf Appendix \chapnum .}
\vskip 0.2truein

Here we prove the important crossing relation
$$t(u) = t(-u -\rho) \,. \eqlabel{important} $$
First we observe that
$$\eqalignno{
t(u) = t(u)^t &= \tr_a \left\{ M_a\ T_a(u)\ \hat T_a(u)
\right\}^{t_1 \cdots t_N} \cr
&= \tr_a \hat T_a(u)^{t_1 \cdots t_N}\ M_a\
T_a(u)^{t_1 \cdots t_N}  \,, \eqalignnum \cr} $$
where the first equality follows from \eqref{symmetric}. But
$$\eqalignno{
T_a(u)^{t_1 \cdots t_N} &= \left\{R_{aN}(u)\ \cdots R_{a1}(u)
\right\}^{t_1 \cdots t_N} \cr
&= R_{aN}(u)^{t_N}\ \cdots R_{a1}(u)^{t_1} \cr
&= V_a\ R_{aN}(-u-\rho)\ V_a\ \cdots V_a\ R_{a1}(-u-\rho)\ V_a \cr
&= V_a\ T_a(-u-\rho)\ V_a  \,, \eqalignnum \cr} $$
and similarly,
$$\hat T_a(u)^{t_1 \cdots t_N} = V_a^{t_a}\ \hat T_a(-u-\rho)\ V_a^{t_a}
\,. \eqnum $$
Thus,
$$\eqalignno{
t(u) &= \tr_a  V_a^{t_a}\ \hat T_a(-u-\rho)\ V_a^{t_a}\
M_a\ V_a\ T_a(-u-\rho)\ V_a  \cr
     &= \tr_a  M_a^{-1}\ \hat T_a(-u-\rho)\ T_a(-u-\rho) \cr
     &= t(-u -\rho) \,, \eqalignnum \cr} $$
where the final step follows from the identity \eqref{useful}.

\vfill\eject

\noindent
{\bf References}

\vskip 0.2truein

\reflabel{alcaraz}
F.C. Alcaraz, M.N. Barber, M.T. Batchelor, R.J. Baxter and G.R.W. Quispel,
J. Phys. {\it A20} (1987) 6397.
See also M. Gaudin, Phys. Rev. {\it A4} (1971) 386;
{\it La fonction d'onde de Bethe} (Masson, 1983).

\reflabel{sklyanin}
E.K. Sklyanin, J. Phys. {\it A21} (1988) 2375.
See also A.B. Zamolodchikov, unpublished;
I.V. Cherednik, Theor. Math. Phys. {\it 61} (1984) 977.

\reflabel{pasquier}
V. Pasquier and H. Saleur, Nucl. Phys. {\it B330} (1990) 523.

\reflabel{kulish/sklyanin(1)}
P.P. Kulish and E.K. Sklyanin, J. Phys. {\it A24} (1991) L435 ; in
{\it Proc. Euler Int. Math. Inst., 1st Semester: Quantum Groups, Autumn 1990},
ed. by P.P. Kulish, in press.

\reflabel{jpa}
L. Mezincescu and R.I. Nepomechie, J. Phys. {\it A24} (1991) L17.

\reflabel{ijmpa}
L. Mezincescu and R.I. Nepomechie, Int'l J. Mod. Phys. {\it A}, in press.

\reflabel{mpla}
L. Mezincescu and R.I. Nepomechie, Mod. Phys. Lett. {\it A6} (1991) 2497.

\reflabel{fusion}
M. Karowski, Nucl. Phys. {\it B153} (1979) 244;
P.P. Kulish, N.Yu. Reshetikhin and E.K. Sklyanin, Lett. Math. Phys. {\it 5}
(1981) 393.

\reflabel{kulish/sklyanin(2)}
P.P. Kulish and E.K. Sklyanin, {\it Lecture Notes in Physics} {\it 151}
(Springer, 1982) 61.

\reflabel{spin1}
L. Mezincescu, R.I. Nepomechie and V. Rittenberg, Phys. Lett. {\it A147}
(1990) 70;
R.I. Nepomechie, in {\it Superstrings and Particle Theory}, ed. by
L. Clavelli and B. Harms (World Scientific, 1990) 319;
L. Mezincescu and R.I. Nepomechie, in {\it Argonne Workshop
on Quantum Groups}, ed. by T.L. Curtright, D. Fairlie and C. Zachos
(World Scientific, 1991) 206.

\reflabel{npb}
L. Mezincescu and R.I. Nepomechie, CERN preprint (1991).

\reflabel{batchelor}
M.T. Batchelor and A. Kuniba, J. Phys. {\it A24} (1991) 2599.

\reflabel{baxter}
R.J. Baxter, Ann. Phys. {\it 70} (1972) 193.

\reflabel{stroganov}
Yu.G. Stroganov, Phys. Lett. {\it 74A} (1979) 116.

\reflabel{zamolodchikov}
A.B. Zamolodchikov and Al.B. Zamolodchikov, Ann. Phys. {\it 120} (1979) 253;
A.B. Zamolodchikov, Sov. Sci. Rev. {\it A2} (1980) 1.

\reflabel{reshetikhin}
V.I. Vichirko and N.Yu. Reshetikhin, Theor. Math. Phys. {\it 56}
(1983) 805; N.Yu. Reshetikhin, Lett. Math. Phys. {\it 7} (1983) 205;
N.Yu. Reshetikhin, Sov. Phys. JETP {\it 57} (1983) 691;
N. Yu. Reshetikhin, Lett. Math. Phys. {\it 14} (1987) 235.

\reflabel{nested}
C.N. Yang, Phys. Rev. Lett. {\it 19} (1967) 1315;
O. Babelon, H.J. de Vega and C-M. Viallet, Nucl. Phys. {\it B200} (1982) 266.

\reflabel{kulish/sklyanin(3)}
P.P. Kulish and E.K. Sklyanin, J. Sov. Math. {\it 19} (1982) 1596.

\reflabel{bazhanov}
V.V. Bazhanov, Phys. Lett. {\it 159B} (1985) 321;
Commun. Math. Phys. {\it 113} (1987) 471.

\reflabel{jimbo}
M. Jimbo, Commun. Math. Phys. {\it 102} (1986) 537;
{\it Lecture Notes in Physics}, Vol. 246 (Springer, 1986) 335.

\reflabel{faddeev}
L.D. Faddeev, N. Yu. Reshetikhin and L.A. Takhtajan, Algebraic Analysis,
{\it 1} (1988) 129;  Leningrad Math. J. {\it 1} (1990) 193.

\reflabel{miami}
L. Mezincescu and R.I. Nepomechie, in {\it Quantum Field Theory,
Statistical Mechanics, Topology and Quantum Groups}, ed. by T.L. Curtright,
L. Mezincescu and R.I. Nepomechie, in press.

\reflabel{gauge}
K. Sogo, Y. Akutsu and T. Abe, Prog. Theor. Phys. {\it 70} (1983) 730.

\reflabel{devega}
C. Destri and H.J. de Vega, LPTHE preprint (1991).

\reflabel{izergin}
A.G. Izergin and V.E. Korepin, Commun. Math. Phys. {\it 79} (1981) 303.

\reflabel{closedXXZ}
H. Bethe, Z. Phys. {\it 71} (1931) 205;
R. Orbach, Phys. Rev. {\it 112} (1958) 309;
C.N. Yang and C.P. Yang, Phys. Rev. {\it 150} (1966) 321, 327;
L.D. Faddeev and L.A. Takhtajan, Russ. Math. Surv. {\it 34} (1979) 11.

\end